\documentclass[twocolumn,english,prl,showpacs,preprintnumbers,superscriptaddress]{revtex4-1}
\usepackage[utf8]{inputenc}
\usepackage{color}
\usepackage{amsmath}
\usepackage{amssymb}
\usepackage{graphicx}
\usepackage{textcomp}
\usepackage{dcolumn}
\usepackage{bm}
\usepackage[right]{eurosym}
\usepackage{float}
\usepackage[english]{babel}
\usepackage{blindtext}
\usepackage{babel}

\begin{document}
	
	\title{Highly efficient double ionization of mixed alkali dimers by intermolecular decay}
	
	\author{A. C. LaForge}
	\email{aaron.laforge@physik.uni-freiburg.de}
	\affiliation{Physikalisches Institut, Universit{\"a}t Freiburg, 79104 Freiburg, Germany}
	\affiliation{Department of Physics, University of Connecticut, Storrs, Connecticut, 06269, USA}
	\author{M. Shcherbinin}
	\affiliation{Department of Physics and Astronomy, Aarhus University, 8000 Aarhus C, Denmark}
	\author{F. Stienkemeier}
	\affiliation{Physikalisches Institut, Universit{\"a}t Freiburg, 79104 Freiburg, Germany}
	\author{R. Richter}
	\affiliation{Elettra-Sincrotrone Trieste, 34149 Basovizza, Trieste, Italy}
	\author{R. Moshammer}
	\author{T. Pfeifer}
	\affiliation{Max-Planck-Institut f{\"u}r Kernphysik, 69117 Heidelberg, Germany}
	\author{M. Mudrich}
	\affiliation{Department of Physics and Astronomy, Aarhus University, 8000 Aarhus C, Denmark}


\begin{abstract}
As opposed to purely molecular systems where electron dynamics proceed only through \textit{intramolecular} processes, weakly-bound complexes like helium droplets offer an environment where local excitations can interact with neighboring embedded molecules leading to new \textit{intermolecular} relaxation mechanisms. Here, we report on a new decay mechanism leading to the double ionization of alkali dimers attached to helium droplets by intermolecular energy transfer. From the electron spectra, the process is similar to the well-known shakeoff mechanism observed in double Auger decay and single photon double ionization~\cite{Carlson1967,Schneider2002}, however, in this case, the process is dominant, occurring with efficiencies equal to, or greater than, single ionization by energy transfer. Although an alkali dimer attached to a helium droplet is a model case, the decay mechanism is relevant for any system where the excitation energy of one constituent exceeds the double ionization potential of another neighboring molecule. The process is, in particular, relevant for biological systems, where radicals and slow electrons are known to cause radiation damage~\cite{Boudaieffa2000}.
\end{abstract}

\flushbottom
\maketitle
%
%
\thispagestyle{empty}


\section*{Introduction}

The correlated action of multiple electrons after photon absorption in atomic and molecular systems has led to the discovery of a variety radiation-induced decay processes (e.g. multiple excitation/ionization or various autoionization channels.)~\cite{Madden1963,Wehlitz1991,Carlson1967}. When the system complexity is increased to larger, more complex systems, new and diverse intermolecular decay mechanisms open up. In particular, processes such as intermolecular Coulombic decay (ICD)~\cite{Cederbaum1997} where energy is exchanged between electronically excited atoms or molecules and their neighbors have been of broad interest; for reviews, see~\cite{Hergenhahn2011,Jahnke2015}. ICD and related intermolecular processes are a potentially important channel for radiation damage of biologically relevant systems~\cite{Gokhberg2014,Trinter2014}. Recently, ICD was measured for the first time in a hydrated biomolecular system~\cite{Ren2018}.

\begin{figure*}
	\begin{center}
		\includegraphics[width=\textwidth]{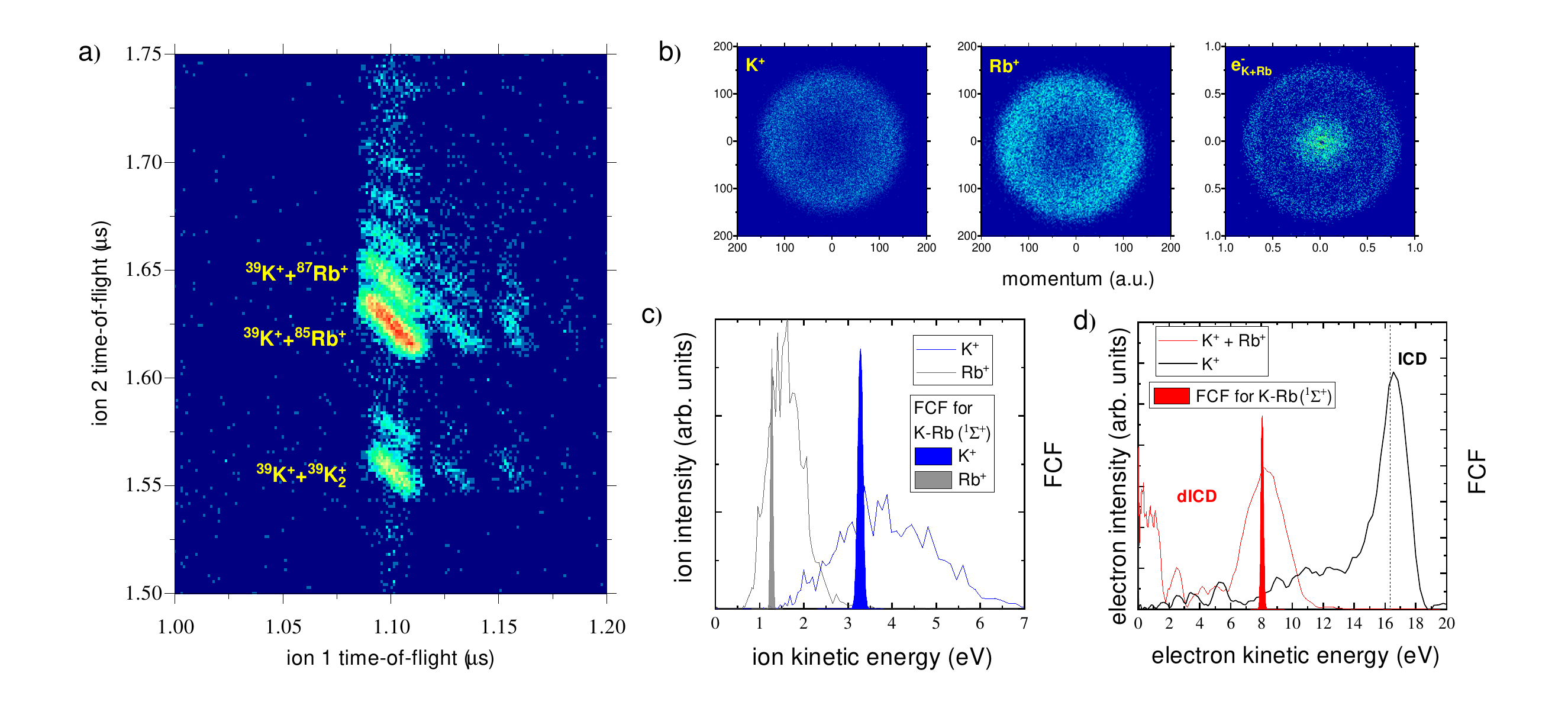}
		\caption{Coincidence spectra for discriminating possible involved decay mechanisms applied while measuring the energetics of the constituent ions and electrons. a) The ion-ion coincidence time-of-flight spectrum for K-Rb dimers attached to the surface of He droplets. The ionization process is triggered by energy transfer from the excited $1s2s^1S$ He atom ($E_e$\,=\,20.6\,eV). b) The K$^+$, Rb$^+$, and electron VMIs taken in triple (e$^-$, $^{39}$K, $^{85}$Rb) coincidence. The momenta scales are given in atomic units. c) The kinetic energy distribution for the K$^+$ (blue line) and Rb$^+$ (gray line) taken in triple (e$^-$, $^{39}$K, $^{85}$Rb) coincidence. d) The electron kinetic energy distributions taken in triple (e$^-$,$^{39}$K, $^{85}$Rb) coincidence (red line) and double (e$^-$, $^{39}$K) coincidence (black line). Note that the black line in d) was a separate measurement where single K atoms were attached to the surface of He droplets. Its expected kinetic energy is given by a dashed vertical line. The filled lines in c) and d) correspond to the respective ion and electron kinetic energy distributions for K-Rb dimers calculated from Franck-Condon factor simulations (see text for details).}
		\label{Fig1}
	\end{center}
\end{figure*}

\begin{figure}
	\begin{center}
		\includegraphics[width=10.5cm]{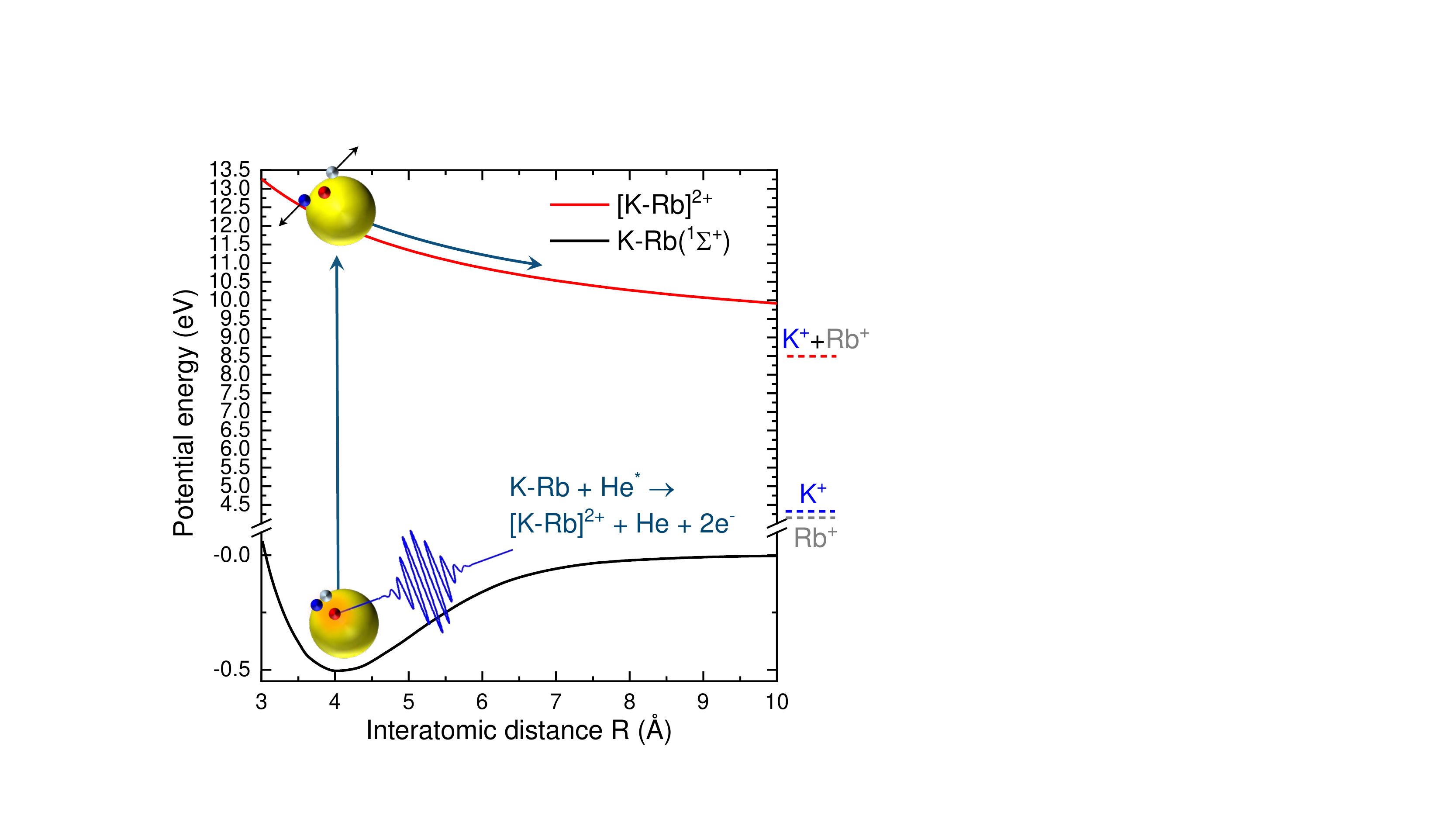}
		\caption{The potential energy curve of K-Rb dimers in the ground (black line) and dicationic (red line) state. The asymptotic limit of the dicationic state is given by the dashed red line along with the individual ionization potentials of K (dashed blue line) and Rb (dashed gray line). A schematic of the process is given where the K-Rb dimer is represented by blue and gray spheres, the He atom by a red sphere, and the He droplet by a yellow sphere. The photon ($h\nu$\,=\,21.6\,eV) is initially absorbed by the He droplet at the $1s2p^1P$ resonance. Through ultrafast intraband relaxation within the droplet~\cite{Kornilov2011,Ziemkiewicz2014}, an excited $1s2s^1S$ He atom ($E_e$\,=\,20.6\,eV) is formed. The excess energy is then transferred by dICD to the K-Rb dimer leading to its double ionization while the He atom relaxes to its ground state.}
		\label{Fig2}
	\end{center}  
\end{figure}

\begin{figure}
	\begin{center}
		\includegraphics[width=10.5cm]{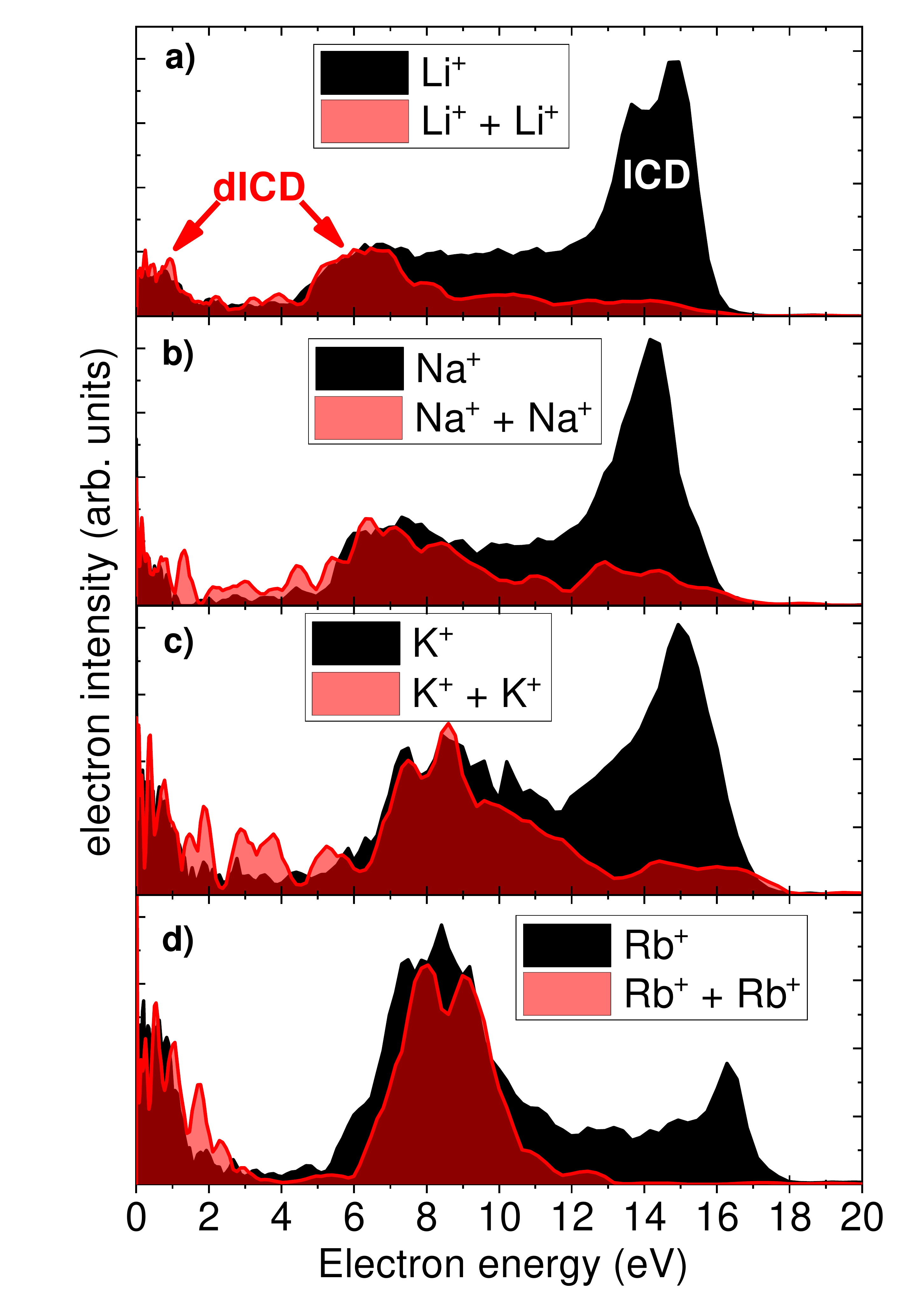}
		\caption{Electron kinetic energy distributions from the ionization of small, homogeneous clusters of alkali metals attached to the surface of a He droplet. Independent spectra for a) Li, b) Na, c) K, and d) Rb. The black filled lines were taken in double (e$^-$, Ak$^+$) coincidence and the red filled lines were taken in triple (e$^-$, Ak$^+$, Ak$^+$) coincidence, where Ak are the alkali metals shown in a)-d). The excited $1s2s^1S$ He atom ($E_e$\,=\,20.6\,eV) triggers the energy transfer process.}
		\label{Fig3}
	\end{center}
\end{figure}


Due to its simple electronic structure and high ionization potential, weakly-bound helium (He) complexes have served as a model system for studying intermolecular processes such as ICD~\cite{Sisourat2010,Havermeier2010a,Shcherbinin2017} and electron transfer mediated decay (ETMD)~\cite{Stumpf2014,Laforge2016}. Additionally, He complexes have been used to observe a related type of ICD. In this case, when He droplets~\cite{Wang2008,Buchta2013} or He-Ne dimers~\cite{Trinter2013} are resonantly excited, the energy can be transferred to neighboring constituents leading to their ionization. An example of such a process is shown in the electron kinetic energy distribution in Fig.\,\ref{Fig1}d) for potassium atoms (K) attached to the surface of He droplets. A photon ($h\nu$\,=\,21.6\,eV) is initially absorbed by a He nanodroplet at the resonance correlating to the $1s2p^1P$-state of atomic He~\cite{Joppien1993}. Through  ultrafast intraband relaxation within the droplet~\cite{Kornilov2011,Ziemkiewicz2014}, an excited $1s2s^1S$ He atom ($E_e$\,=\,20.6\,eV) is formed. The excess energy is then transferred by ICD to the K atom leading to its ionization while the He atom relaxes to its ground state. The characteristic electron kinetic energy is the difference between the He excited state and the acceptor's ionization potential. For the case of K atoms, this results in a kinetic energy of about 16.3\,eV, which matches the position of the pronounced peak in the spectrum (black line in Fig.\,\ref{Fig1}d)). A question which arises is how the situation changes for systems where double ionization is energetically allowed. For endofullerenes, it has been theoretically proposed that double ICD (dICD) can become a viable decay mechanism~\cite{Averbukh2006}. Here, we show that, indeed, dICD is not only a possible decay path, but can even be the dominant decay mechanism occurring with efficiencies equal to, if not exceeding, single ionization.

\section*{Results and Discussion}

To gain detailed insight into the process, we turn our investigation to alkali dimers, the simplest metal cluster. Furthermore, to circumvent issues with detector dead times, the dimers are composed of alkali atoms of different mass, K and Rb (rubidium). The process of dICD is schematically shown in Fig.\,\ref{Fig2} along with the potential energy curves of free K-Rb dimers in the ground $X^1\Sigma^+$ state (black line)~\cite{Rousseau2000} and dicationic state (red line). The dicationic curve was calculated using a Coulomb potential shifted to match the asymptotic ionization energy of the free atoms, which are given as dashed lines in Fig.\,\ref{Fig2}. Similar to the ICD ionization of K atoms described above, dICD occurs by a transfer of energy from the excited $1s2s^1S$ He atom ($E_e$\,=\,20.6\,eV) to the K-Rb dimer. However, in this case, the double ionization potential of the K-Rb dimer is energetically less than the excited He atom resulting in the emission of two electrons along with the dicationic dissociation of the ions. 

Direct evidence for dICD is thus determined by measuring multiple coincidences of electrons and ions produced by this process. Fig.\,\ref{Fig1} a) shows the electron-ion-ion coincidence time-of-flight spectrum for K-Rb dimers attached to the surface of He droplets for a photon energy of 21.6\,eV. In general, distributions observed in ion-ion coincidence maps identify ions created by multiple ionization while the shape of the distribution gives information about the dissociation process~\cite{Eland1991}. In this case, the coincidence map is centered around the respective masses of K and Rb where several sharp, negative sloping features are observed. These distributions indicate that fragmentation occurs through dicationic dissociation of the dimers leading to back-to-back emission of the ions. The primary ion pair originates from dimers of $^{39}$K and $^{85}$Rb while the neighboring distributions come from the isotopes, $^{41}$K and $^{87}$Rb. There are additional, weaker distributions due to complexes of an alkali ion with a few He atoms attached. For cases where the He droplet is not resonantly excited, no such distributions are observed indicating that ionization proceeds through excited He atoms. Additionally, using electron-ion-ion coincidence imaging techniques, one can extract electron/ion kinetic energy spectra from the individual ion pairs in the coincidence map (see Fig.\,\ref{Fig1} c) and d)). 

Fig.\,\ref{Fig1} b) shows the raw velocity map images (VMIs) of ions and electrons measured in triple (e$^-$, $^{39}$K, $^{85}$Rb) coincidence. VMIs are two-dimensional projections of the charged particle's momentum sphere which are then inverted to obtain kinetic energy distributions for the respective electrons and ions~\cite{Dick2014}. The left and middle images show VMIs of K$^+$ and Rb$^+$ ions, and the right image shows the electron VMI. The clearly visible ring structure in all VMIs indicates a non-zero kinetic energy component. 

From the VMIs, we determine the kinetic energy distributions by Abel inversion~\cite{Dick2014}. Fig.\,\ref{Fig1} c) shows the ion kinetic energy distributions for the $^{39}$K ion (blue line) and $^{85}$Rb ion (gray line) measured in triple (e$^-$, $^{39}$K, $^{85}$Rb) coincidence. The ions have broad kinetic energies centered around 3.75\,eV and 1.5\,eV, respectively. The sum of these energies corresponds to the kinetic energy release of the ion pair in the dicationic state as illustrated in Fig.\,\ref{Fig2}. To assess this conjecture, we have performed Franck-Condon factor (FCF) simulations of the ion and electron kinetic energy distributions assuming vertical transitions between the potential energy curves given in Fig.\,\ref{Fig2}. The initial state is assumed to be an excited He atom in the $1s2s^1S$-state ($E_e$\,=\,20.6\,eV) interacting with the alkali dimer in its vibronic ground state.~\cite{Wang2008,Buchta2013}. The results are shown as filled peaks. Note that the $1s2s^1S$-state of a He atom in a droplet is still dipole-coupled to the ground state~\cite{Joppien1993}, thereby allowing ICD-like energy transfer to occur. The kinetic energy release from the FCF simulations, shown in Fig.\,\ref{Fig1} c), gives quantitatively similar results to the measured values, but drastically underestimates the width. Broadening of the experimental distributions is likely due to perturbations of the initial and final, ionic state by dopant-He droplet interactions. In particular, the transient attachment of the localized excited He atom to the K-Rb dimer may lead to its stabilization. Depending on the configuration of the state, dICD proceeds at different internuclear distances of the K-Rb dimer resulting in a broader distribution of the fragmented ions.

Fig.\,\ref{Fig1} d) shows the electron kinetic energy distribution (red line) measured in triple (e$^-$, $^{39}$K, $^{85}$Rb) coincidence. The spectrum shows two peaks centered at 0\,eV and 8\,eV, which arise from double ionization of alkali dimers. The simulated excess electron energy for double ionization of K-Rb dimers is 8\,eV (filled peak) fitting well with the sum electron energy. The measured kinetic energy spectrum shows a U-shaped distribution indicating one electron takes nearly all of the excess energy while the second electron is emitted with nearly zero kinetic energy. Similar distributions have previously been observed in single-photon double ionization of atoms (SPDI) ~\cite{Wehlitz1991,Schneider2002} and double Auger decay (DAD)~\cite{Carlson1967,Viefhaus2004}. In those cases, the mechanism, known as shakeoff, is due to the sudden removal of the primary electron leaving the system in a perturbed ionic state; the secondary electron then has a probability of relaxing to an unbound state resulting in an unequal sharing of the excess energy. The electron energy distribution in Fig.\,\ref{Fig1} d) shows a similar distribution to shakeoff, but, in contrast, occurs relatively close to the double ionization threshold. This could, in part, be due to the low ionization potential of the valence electron for alkali atoms. The overall similarity to shakeoff indicates that dICD proceeds through a one-step process as opposed to other two-step electron impact ionization mechanisms in SPDI such as knockout~\cite{Schneider2002}.

We have verified dICD for several mixed alkali dimer systems (K-Rb, Na-K, Na-Rb), small homogeneous alkali clusters (Li, Na, K, Rb, and Cs), and even alkaline earth atoms (Ba). Fig.\,\ref{Fig3} shows the electron kinetic energy distributions of small, homogeneous clusters of a) Li, b) Na, c) K, and d) Rb attached to the surface of a He droplet. The excited $1s2s^1S$ He atom ($E_e$\,=\,20.6\,eV) triggers the energy transfer process. The black filled lines were taken in double (e$^-$, Ak$^+$) coincidence and the red filled lines were taken in triple (e$^-$, Ak$^+$, Ak$^+$) coincidence with the alkali metal ions, Ak$^+$, where Ak denotes Li, Na, K, Rb. The red filled lines show electrons emitted from dICD while the black filled lines show electrons emitted from ICD, occurring at higher kinetic energies, as well as dICD. Due to the comparable electronic structure and ionization potentials of alkali metals, their distributions exhibit similar features. In all cases shown, dICD is a prominent decay channel leading one to conclude the process is rather ubiquitous and not limited specifically to K-Rb dimers where the excited ionic state of Rb could also lead to double ionization through a cascade mechanism. In particular, the asymmetric distribution observed in dICD is evident of a similar one-step, shakeoff-type ionization mechanism.

Surprisingly, as can be seen in Fig.\,\ref{Fig3}, dICD is a highly efficient process showing comparable, if not larger, ionization rates to ICD. In contrast, for SPDI near threshold, the branching ratio to single ionization is much less than 1\% for atoms~\cite{Samson1998} and small molecules~\cite{Masuoka1994}. For DAD, the branching ratio to Auger decay is typically a few percent for atoms~\cite{Kolorenc2016}. As such, one can conclude that dICD can even be the dominant process in weakly-bound systems for cases where it is energetically allowed. In general, dICD should not be limited to outer valence shell excited atoms; Auger forbidden inner valence shell excited/ionized atoms, which have even higher excitation energies, have the potential for dICD as well. Additionally, the multiple ions and electrons formed in the process of dICD should play an important role in biological systems. For instance, core shell-ionization of a solvated magnesium dication leads to a variety of cascade channels where Auger and intermolecular decay processes occur~\cite{Stumpf2016}. For each step where ICD is allowed, dICD would also be an energetically open decay channel leading to an enhancement in the production of neighboring water ions and low energy electrons. The subsequent ionization of water typically leads to proton transfer and the formation of the hydroxyl radical, a highly reactive damage center~\cite{Jonah2001}, while the production of low energy electrons is a known source of radiation damage for proteins and DNA~\cite{Boudaieffa2000}. 

\section*{Methods}
\subsection*{Experimental setup}
The experiment was performed using a mobile He droplet machine attached to a velocity map imaging photoelectron photoion coincidence spectrometer~\cite{OKeeffe2011} at the GasPhase beamline of Elettra-Sincrotrone Trieste, Italy. The setup has been described in some detail earlier~\cite{Buchta2013}, and only the significant points will be addressed here. In short, a beam of He nanodroplets is produced by continuously expanding pressurized He (50 bar) of high purity (He 6.0) out of a cold nozzle (T = 7 $-$ 40 K) with a diameter of 5 $\mu$m into vacuum. Under these expansion conditions, the mean droplet sizes range from 10$^1$ to 10$^{8}$ He atoms per droplet~\cite{Toennies2004}. After passing a skimmer (0.4 mm) and a mechanical beam chopper used for discriminating the droplet beam signal from the He background, the droplets were doped using the “pick-up” technique~\cite{Gough1985} with subsequent heated doping cells filled with alkali metals. While most atomic and molecular species become submerged into the interior of He nanodroplets, alkali atoms remain weakly-bound on the surface~\cite{Barranco2006}. The He droplet beam next crosses the synchrotron beam inside of a PEPICO detector consisting of an ion time-of-flight detector and velocity map imaging detector (5\% $\Delta E/E$ resolution). With this setup, one can record either electron or ion kinetic energy distributions, depending on the polarity, in coincidence with one specific ion mass or with several ion masses in multicoincidence mode~\cite{OKeeffe2011}. When electrons are recorded on the VMI, only one electron for each coincidence event can be detected.  The kinetic energy distributions were reconstructed using the Maximum Entropy Legendre Reconstruction method~\cite{Dick2014}. The polarization axis was perpendicular to the VMI axis to ensure cylindrical symmetry which is required for the inversion process. The photon energy was set to 21.6\,eV and a tin filter was used to eliminate any higher-order light contamination. The pulse repetition rate was 500 MHz.

\section*{Data availability}
The data that support the plots within this paper and other findings of this study are available from the corresponding authors on request.

\section*{Author contributions statement}
A.C.L. and M.M. conceived the experiment. A.C.L, M.S., and R.R. conducted the experiment. A.C.L, M.S., and M.M. analyzed the data. M.M. performed the FCF calculations. A.C.L. interpreted the data with help from R.R., F.S., R.M., T.P., and M.M.. A.C.L. wrote the paper. All authors reviewed the manuscript.

\end{document}